\documentclass[twoside,final]{pro12}
\usepackage[ansinew]{inputenc}
\usepackage{amsmath}
\usepackage{graphicx} 

\head{D. E. Morosan and P. T. Gallagher} {Characteristics type III Radio Bursts and Solar S bursts} \label{Morosanpaper}
\begin{document}

\title{Characteristics of type III Radio Bursts and Solar S bursts}
\author{D. E. Morosan\adress{School of Physics, Trinity College Dublin, Dublin 2, Ireland } ~and P. T. Gallagher$^*$$\,$}


\maketitle \thispagestyle{empty}

\begin{abstract}

The Sun is an active source of radio emission which is often associated with the acceleration of electrons arising from processes such as solar flares and coronal mass ejections (CMEs). At low radio frequencies ($<$100~MHz), numerous solar S bursts (where S stands for short) and storms of type III radio bursts have been observed, that are not directly relates to flares and CMEs. Here, we expand our understanding on the spectral characteristic of these two different types of radio bursts based on observations from the Low Frequency Array (LOFAR). On 9 July 2013, over 3000 solar S bursts accompanied by over 800 type III radio bursts were observed over a time period of $\sim$8 hours. The characteristics of type III radio bursts presented here are consistent with previous studies. S bursts are shown to be different compared to type III bursts: they show narrow bandwidths, short durations and drift rates of about 1/2 the drift rate of type III bursts. Both type III bursts and solar S bursts occur in a region in the corona where plasma emission is the dominant emission mechanism as determined by data constrained density and magnetic field models.

\end{abstract}

\section{Introduction}


{Solar activity is often accompanied by numerous solar radio bursts such as the type I, II, III, IV and V [Wild et al. 1950a, 1950b, 1950c, Boischot 1957, Wild et al. 1959] but also numerous other radio bursts which exhibit a variety of fine frequency structures and time scales of $<$1~s in dynamic spectra, an example being solar S bursts [McConnell 1982, Morosan et al. 2015]. At frequencies $<$100~MHz where numerous type III radio bursts and S bursts occur, there have been only a number of studies due to sensitivity limitations and the low spectral and temporal resolution of previous observations. }

{Type IIIs are rapidly varying bursts of radiation at metre wavelengths with durations of a few seconds. They were first described by Wild [1950] and they were found to have a characteristic frequency drift from high to low frequencies in solar dynamic spectra. Since then, type IIIs have been observed at kHz frequencies [Krupar et al. 2010] up to frequencies of 8~GHz [Ma et al. 2012], although most occur at frequencies $<$150~MHz [Saint--Hilaire et al. 2013]. Their drift rates observed in dynamic spectra vary from approximately -1~MHz s$^{-1}$ at 20~MHz to -20~MHz s$^{-1}$ at 100~MHz [Abranin et al. 1990, Mann and Klassen, 2002], while their brightness temperatures can be up to $10^{12}$~K for coronal type IIIs. High brightness temperature type IIIs are generally associated with flaring activity on the Sun, however 90\% of the time type III bursts occur in the absence of flares or CMEs [Dulk 1985].}

{Type III bursts are considered to be the radio signature of electron beams travelling through the corona and into interplanetary space along open magnetic field lines [Lin 1974] and these electron beams are believed to originate from magnetic reconnection or shocks [Dulk et al. 2000]. There are a number of theoretical explanations for type III bursts, but it is commonly believed that, following acceleration, faster electrons outpace the slower ones to produce a bump-on-tail instability in their velocity distribution. This generates Langmuir (plasma) waves [Robinson et al. 1993] which are then converted into radio waves at the local plasma frequency ($\omega_p$) and its harmonic ($2\omega_p$, Bastian et al. 1998).}

{S bursts were first identified by Ellis [1969] as a new type of short duration radio bursts which were initially named fast drift storm bursts. McConnell [1982] also observed these bursts and he showed that their drift rate is slower than that of type III radio bursts (about 1/3 the drift rate of type IIIs) which are a form of plasma emission. McConnell [1982] renamed fast drift storms to ``solar S bursts" due to their similarity to jovian S bursts.}

{S bursts appear as narrow short drifting lines in dynamic spectra with a total duration of about 1 s. They have been observed in a frequency range of 10--150~MHz [Ellis 1982; McConnell 1982; Melnik et al. 2010; Morosan et al. 2015]. They have short instantaneous bandwidths of about 120 kHz and short FWHM duration of 50 ms at frequencies of 40 MHz [McConnell 1982]. Most S bursts have negative drift rates, i.e. from high to low frequencies, however a few S bursts have positive drift rates [McConnell 1982, Briand et al. 2008, Morosan et al. 2015]. All S bursts occurred during times of other solar activity such as type III and type IIIb radio bursts [McConnell 1982, Briand et al. 2008, Melnik et al. 2010]. }

{So far, plasma emission has been proposed by Zaitsev and Zlotnik [1986] and Melnik et al. [2010] as the mechanism responsible for the emission of solar S bursts. Zaitsev et al. [1986] suggested that electrons with velocities 10--20 times above their thermal velocity excite plasma waves near the upper hybrid resonance frequency. These waves are then scattered by ions producing electromagnetic waves at the plasma frequency ($\omega_p$). Melnik et al. [2010] propose a model in which S bursts are generated by the coalescence of magnetosonic waves and Langmuir waves and derive a minimum magnetic field of 2~G necessary for the generation of these bursts at the plasma frequency ($\omega_p$). However S bursts have a significantly slower drift rate than type III radio bursts (about 1/3 of the type III drift rate), very narrow bandwidths and short lifetimes and they are significantly more polarised. The combination of narrow bandwidth, short duration and high degree of polarisation is indicative of electron-cyclotron maser (ECM) emission. Solar S bursts are very similar in appearance to jovian S bursts which are an example of ECM emission generated by $\sim5$~keV electrons accelerated in the flux tubes connecting Io to Jupiter [Zarka et al. 1996]. These electrons have an adiabatic motion along the magnetic field lines and they are magnetically mirrored at the foot of the Io flux tube. Near Jupiter they emit radio bursts triggered by the loss cone instability in the magnetically mirrored population of electrons. Hess et al. [2007] simulated that these electrons are accelerated by Alfv\'en waves which generate electron beams that will travel along the Io-Jupiter flux tube to generate ECM emission. The nature of S bursts as very discrete bursts is explained by the Alfv\'en wave acceleration scenario where electrons are accelerated in a very narrow region named the Alfv\'en wave resonator where short-lived electron beams are produced [Hess et al. 2007, Hess et al. 2009].}

{With the aid of the Low Frequency Array (LOFAR; van Haarlem et al. 2013) over 800 type III radio bursts and over 3000 solar S bursts have been observed on 9 July 2013 in high time and frequency resolution dynamic spectra. In this paper, LOFAR tied-array beam observations were used to study the spectral characteristics of type III bursts and S bursts in order to compare these bursts and identify a possible emission mechanism of S bursts.}

\section{Observations}

{LOFAR is a next-generation radio interferometric array constructed by the Netherlands Institute for Radio Astronomy (ASTRON) to operate at low radio frequencies (10--240~MHz). It consists of two antenna types:  Low Band Antennas (LBAs) operating at frequencies of 10--90~MHz and High Band Antennas (HBAs) which operate at 110--240~MHz [van Haarlem 2013]. There are over 7000 antennas distributed in 24 core stations and 14 remote stations across the Netherlands and 12 international stations across Europe. }

{In this study, we used LOFAR tied-array beam observations of the Sun [Stappers et al. 2011, van Haarlem et al. 2013] in the LBA frequency range. We used 170 simultaneous beams to cover a field-of-view of 1.3$^{\circ}$~centred on the Sun produced by the LOFAR core stations. An example of pattern of the tied array beams is shown in Figure 1a. Each beam produces a high time and frequency resolution dynamic spectra ($\sim$10~ms; 12.5~kHz) at a unique spatial location as shown in Figure 1b and 1c. The dynamic spectra can be used to produce tied-array images of radio bursts that can give an approximate position of radio sources (for more details, see Morosan et al. 2014, 2015). }

{On 9 July 2013 starting at 07:00 UT, over 3000 S bursts and over 800 type III radio bursts were observed during an 8 hour LOFAR observation campaign [Morosan et al. 2015]. At higher frequencies ($>$150~MHz), a type I storm was observed by the Nan{\c c}ay Radioheliograph above the active region NOAA 11785, which is a complex $\beta\gamma\delta$ region. Figure 1 shows an example of the S bursts and type III radio bursts observed over a time period of 60~s occurring concurrently but corresponding to dynamic spectra recorded at different locations. Due to a variety of shapes (various lengths, time profiles, frequency profiles) and the fact that S bursts are similar in intensity to type III/type IIIb bursts, it was necessary to identify these bursts manually to extract their spectral characteristics.  }

\begin{figure}[ht]
\includegraphics[width = 230 px, angle = -90, trim = 160px 85px 100px 40px]{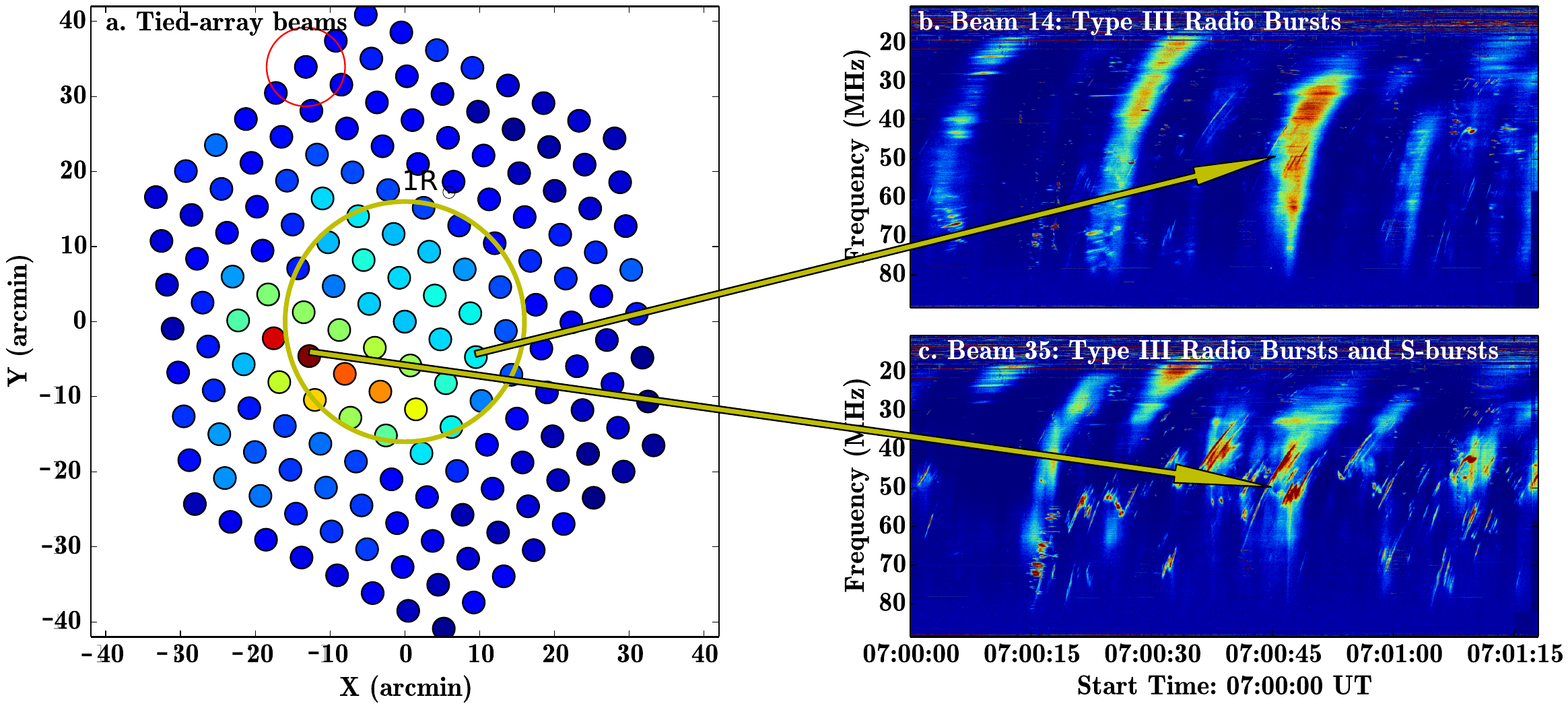}
\caption{LOFAR tied-array beam observations of type III radio bursts and solar S bursts [Morosan et al. 2015]. (a) Map of 170 tied-array beams covering a field-of-view of $\sim$1.3$^{\circ}$about the Sun. The full-width half-maximum (FWHM) of the beams at zenith at a frequency of 60~MHz is represented by the red circle and the size of the optical Sun is represented by the yellow circle. (b) Dynamic spectrum recorded for a period of 1.3 minutes corresponding to Beam 14 (west) containing type III radio bursts. (c) Dynamic spectrum recorded for a period of 1.3 minutes corresponding to Beam 35 (east) containing type III radio bursts and S bursts. The yellow arrows indicate the locations where the two dynamic spectra were recorded in (b) and (c) pointing at the time and frequency corresponding to the intensity values in (a). \label{fig1}}
\end{figure}

\section{Results}

{An example of type III radio bursts and solar S bursts occurring concurrently on the Sun is shown in Figure 1 where two dynamic spectra recorded solar activity at two separate locations (east and west) denoted by the yellow arrows [Morosan et al. 2015]. Type III bursts are the wide long-bandwidth bursts in Figure 1a, while solar S bursts are the narrow sloping lines in Figure 1b. The characteristics of these bursts such as bandwidth, duration, instantaneous bandwidth and the duration at a fixed frequency or the full width half max (FWHM) duration were estimated as shown in Figure 2. Due to a variety of shapes and intensities and due the the fact that S-burst and type IIIs overlapped in dynamic spectra, it was necessary to identify these bursts manually in order to determine their spectral properties. In particular, the start and end points of the bursts were identified for total duration, bandwidth and drift rate estimations and the FWHM was estimated for a subset of these bursts in order to determine instantaneous bandwidth and FWHM duration. The estimated spectral properties of type III and S-bursts are shown in Table 1. }		

{Type IIIs where observed over the entire frequency band of the LBA up to 80~MHz where the signal is significantly suppressed due to radio frequency interference filters. Their duration varied from 1~s for narrowband type IIIb bursts to $\sim$15~s for broader and diffuse type III bursts. Their bandwidth also varied from 10~MHz up to the entire frequency band of the LBA. The frequency drift rates of type III bursts varied between -1 to -20~MHz~s$^{-1}$ with an average of -7.4~MHz~s$^{-1}$. This drift rate is slower than the findings of Mann and Klassen [2002], where a mean value of -11~MHz~s$^{-1}$ was reported at a frequency range of 40--70~MHz.}

\begin{figure}[t]
\centering
\includegraphics[width = 250px, angle = 90, trim = 80px 0px 55px 0px ]{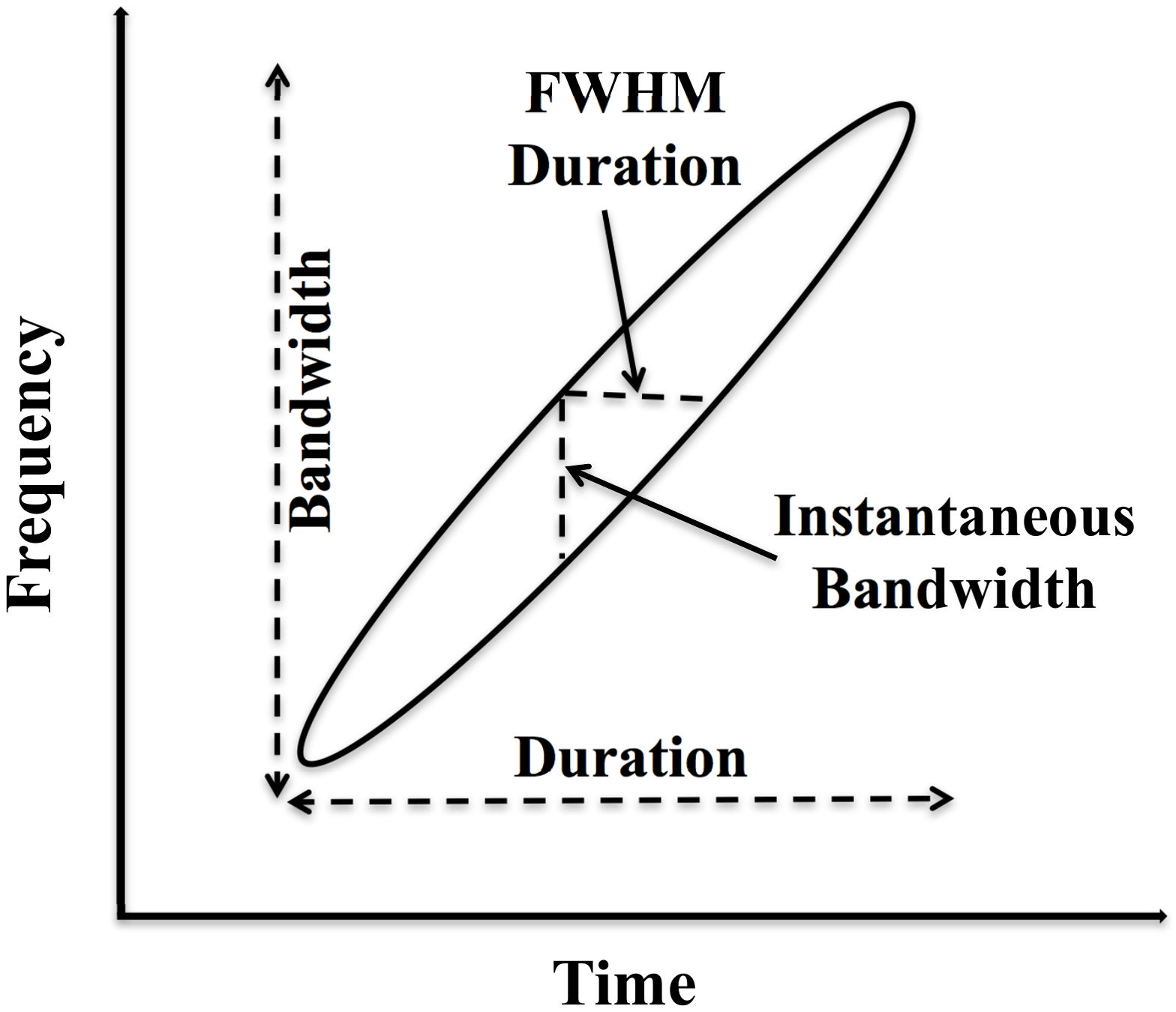}
\caption{Idealised diagram of a solar S burst showing the properties that can be extracted from the dynamic spectrum: duration, bandwidth, FWHM duration and instantaneous bandwidth. }
\end{figure}

\begin{table}[ht]
\centering
	\begin{tabular}{lcc} 
	\hline\hline
	 & S bursts & type III bursts\\ \hline \hline
	Frequency range (MHz) & 20--80 & 10--80 \\ 
	Duration (s) & 0.2--1 & 1--15\\ 
	FWHM duration (s) & 0.02--0.4 & 0.5--5 \\ 
	Total bandwidth (MHz) & 1--5 & 10--70 \\ 
	Instantaneous bandwidth (MHz) & 0.1--1.5 & 5--40\\ 
	Drift rate (MHz s$^{-1}$) & -1 to -7  & -1 to -20 \\ \hline \hline
	\end{tabular}
	\caption{\label{table1}Typical S bursts and type III bursts characteristics estimated from LOFAR dynamic spectra in the LBA frequency range.} 
\end{table}

{Most S bursts had short durations of $<$1~s and FWHM durations at a fixed frequency of $<$400~ms, some as small as 20~ms. They were observed at frequencies between 20 to 80~MHz and the majority had a total bandwidth of $\sim$2.5~MHz and instantaneous bandwidths at a fixed time varying between 0.1--1.5~MHz. Very few S bursts had larger bandwidths $>$20~MHz. Their drift rates varied with frequency from -1~MHz~s$^{-1}$ at frequencies of 20~MHz to -7~MHz~s$^{-1}$ at 75~MHz but most bursts had a drift rate of -3.5~MHz s$^{-1}$. The typical S burst drift rate is about 1/2 of the type III drift rate since the average type III drift rate is -7.4~MHz s$^{-1}$ in these observations. }

{Frequency drift rates calculated for all S bursts and type III radio bursts observed on 9 July 2013 are shown in Figure 4. The drift rates for both S bursts (red circles) and type III radio bursts (green triangles) are plotted as a function of centre frequency. Figure 4 shows that most S bursts have a lower drift rate than type III radio bursts and a clear increase in drift rate with increasing frequency as observed before by McConnell [1982]. The cyan curve represents the fit proposed by McConnell [1982] in the case of S bursts and Mann et al. [1999] in the case of type III bursts. McConnell [1982] found the following relation between drift rate and frequency for S bursts: 
\begin{align}
\label{eq5-1}
\frac{df}{dt} = - af^b
\end{align} 
where $a = 6.5\times10^{-3}$ and $b = 1.60$. The cyan curve in Figure 4 fits the LOFAR data very well except for a few high drift rate S bursts at frequencies between 50--80~MHz. The blue curve is a more accurate fit to the LOFAR data where the fitting parameters are: $a = 4.9 \times10^{-3}  \pm1.1\times10^{-3}  $ and $b = 1.70 \pm 0.05$ which takes into account the S bursts drift rates that were observed to be spread at higher frequencies. The blue curve is a better determined fit since over 3000 data points were used as opposed to only a few S bursts analysed by McConnell [1982]. Melnik et al. [2010] also calculated these parameters for S bursts at frequencies between 10--30~MHz and the fitting coefficients were found to be:  $a = 13 \times10^{-3} $ and $b = 1.4$. These coefficients differ from those obtained by McConnell [1982] and the fit to the LOFAR S-bursts, most likely because S-bursts were observed over a short bandwidth between 10--30~MHz.}

{The purple and magenta lines in the plot show the fit to type III radio bursts where the fitting parameters were taken from Mann et al. [1999] (where $a = 7.4\times10^{-3}$ and $b = 1.76$) and the fit to the LOFAR data (where $a = 6.8\times10^{-3}  \pm 3.0 \times10^{-3}$ and $b = 1.82 \pm 0.11$), respectively. Alvarez and Haddock [1973] also calculated these parameters for type IIIs and obtained similar coefficients to those of the LOFAR Type III fit: $a = 10.9 \times10^{-3}  \pm1.1\times10^{-3}  $ and $b = 1.84 \pm 0.02$. The parameters for the type III fit are very close in magnitude to those of the LOFAR S bursts fit suggesting that a similar mechanism is generating these bursts. }

\begin{figure}[t]
\centering
\includegraphics[width = 420px, trim = 0px 0px 0px 10px ]{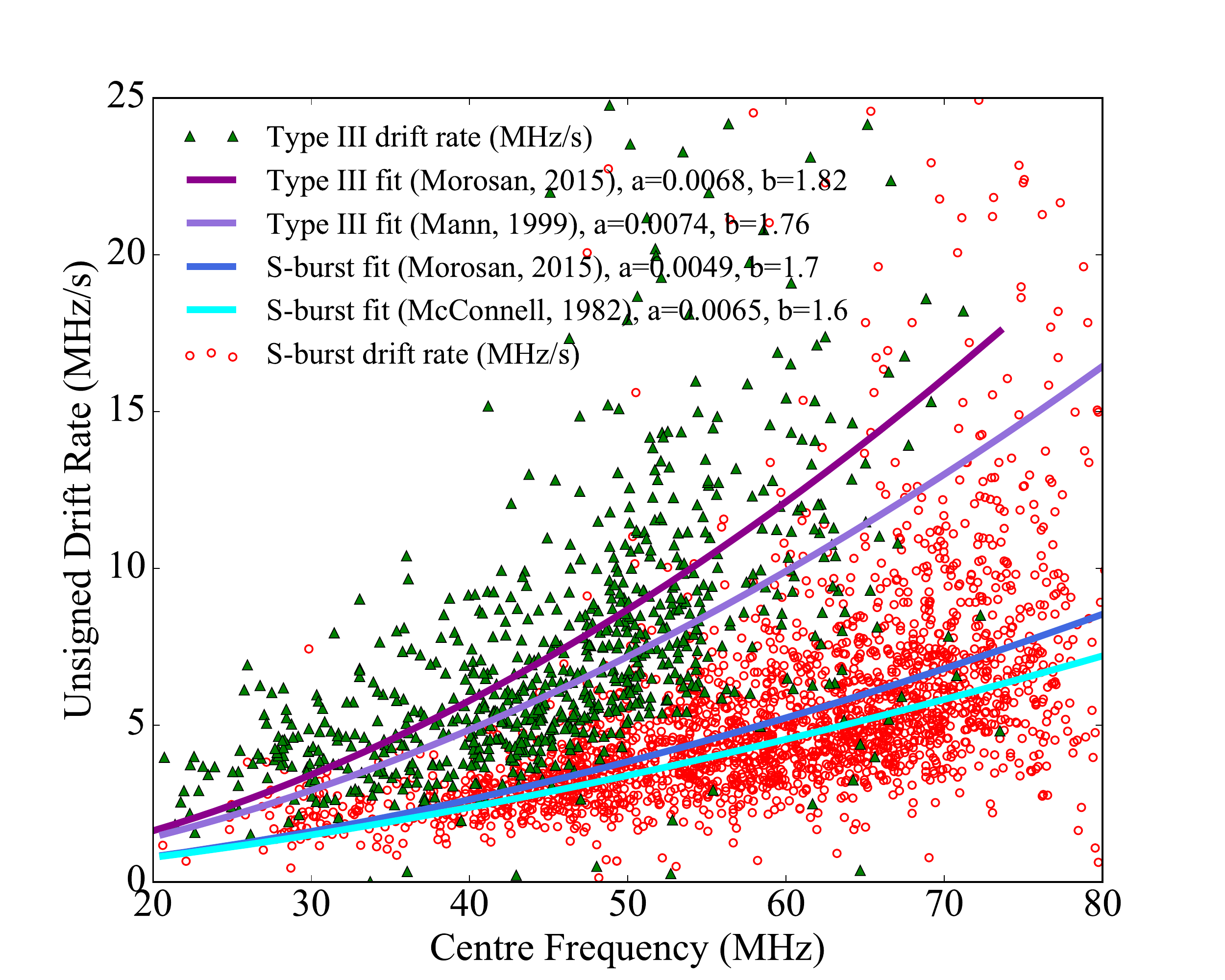}
\caption{Drift rate as a function of frequency for solar S bursts (red circles) and type III radio bursts (green triangles) observed on 9 July 2013. The cyan and blue lines represent the fitting of S bursts data by McConnell [1982] and the best fit of the LOFAR S bursts data, respectively. The purple and magenta lines represent the fitting of the type III radio burst data using the fitting parameters from Mann et al. [1999] and the best fit of the LOFAR type III bursts data, respectively. }
\end{figure}

{In the case of Jovian S bursts, electrons travel adiabatically along a flux tube and generate S bursts as ECM emission after being magnetically mirrored at the Io flux tube foot point. From adiabatic theory, the drift rate of these bursts increases with frequency up to a maximum drift rate around 25~MHz~s$^{-1}$ where the acceleration of the electrons travelling upwards from the mirror point competes with the magnetic field decrease over distance [Zarka et al. 1996]. This behaviour is not observed in solar S bursts and instead the drift rate continues to increase with frequency as with type III bursts which suggests that plasma emission may be responsible for the generation of S bursts as well. }

{In order to determine if S bursts are indeed not emitted by ECM emission based on the analogy with jovian S bursts, the likelihood of ECM emission on the Sun has been investigated. One of conditions for ECM to occur is when the electron-cyclotron frequency $\Omega_e$ is greater than the plasma frequency $\omega_p$ [Melrose 1991]. The ratio of $\omega_p$/$\Omega_e$ dictates if ECM is likely to occur and it can be calculated using data-constrained density and magnetic field models of the solar corona. In order to satisfy the condition $ \omega_p < \Omega_e$ in the solar corona, a high $B$ and a low $n_e$ is required at the emission site, since: 
\begin{equation}
\omega_p = \sqrt{\frac{n_e e^2}{m_e\epsilon_0}}~~ ,
\end{equation}
and,
\begin{equation}
\Omega_e= \frac{eB}{m_e}~~ , 
\end{equation} 
where $n_e$ is the electron plasma density, $B$ is the magnetic field strength and the remaining quantities are known physical constants. A high magnetic field can only be found in active regions.}

\begin{figure}[t]
\centering
\includegraphics[width = 510px, trim = 100px 0px 0px 0px ]{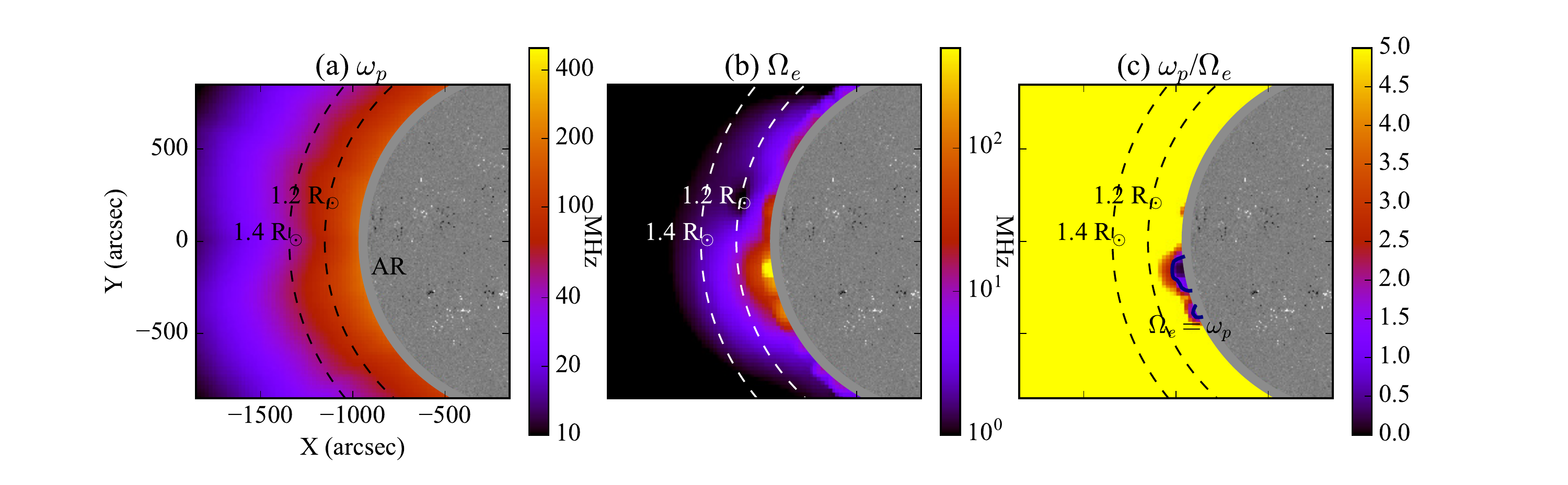}
\caption{Characteristics of the solar corona on 1 July 2013 when the $\beta\gamma\delta$ active region was located on the solar limb: (a) plasma frequency, (b) electron-cyclotron frequency and, (c) plasma frequency to electron-cyclotron frequency ratio [Morosan et al. 2016]. The location of the active region NOAA 11785 on the limb (AR) is labelled in panel (a).}
\end{figure}

{The active region NOAA 11785 was the largest active region on 9 July 2013 on the visible solar disc and it was part of a complex group of four $\beta\gamma\delta$ active regions. This active region group was shown to be associated with the occurrence of solar S bursts [Morosan et al. 2015]. We constructed a map of the electron densities to determine $\omega_p$ (Equation 2) and a map of the magnetic field strength to determine $\Omega_e$ (Equation 3) above this active region in order to determine if ECM emission was likely to occur instead of plasma emission [Morosan et al. 2016]. We used the method of Zucca et al. [2014] to calculate the off-limb electron densities in the corona up to a height of 2.5~solar radii (R$_{\odot}$), where 1~R$_{\odot}$ is 695~Mm (see Zucca et al. 2014 for further details). The electron densities were estimated on 1 July 2013 when NOAA 11785 was located on the eastern solar limb (Figure 4a) so that the plane-of-sky densities are centred on the Carrington longitude of the active region. } 

{The magnetic field was estimated using a potential field source surface (PFSS) model that provides an approximation of the coronal magnetic field at heights up to 2.5~R$_\odot$ based on the observed photospheric field [Schrijver and De Rosa, 2003]. We used the PFSS solution from 4 July 2013 when the active region was not as close to the limb to extract the longitudinal slice through the active region that is equivalent to the Earth-viewed plane-of-sky on 1 July 2013 [Morosan 2016]. This was necessary as the photospheric magnetic field is only accurately represented in the PFSS bottom boundary when the active region is not located close to the east limb (owing to the transition from the flux-transported surface field on the unobserved hemisphere to the observed field on the Earth-facing hemisphere). The magnetic field structures above the Carrington longitude of the active region remained consistent throughout a period of seven consecutive days thereafter, so no major changes occurred before the S bursts observations.}

{Using Equations (2) and, (3), we created maps of $\omega_p$ (Figure 4a), $\Omega_e$ (Figure 4b), and the ratio $\omega_p/\Omega_e$ (Figure 4c). Figure 4c shows that the locations where $\omega_p < \Omega_e$ occurs are at heights below 1.1~R$_\odot$ from the centre in the active region. ECM emission is a possible emission mechanism in the solar corona, but only at heights  below 1.1~R$_\odot$ in the active region studied. These results agree with R\'egnier et al. [2015], where estimates of the $\omega_p/\Omega_e$ ratio showed that ECM is possible at heights up to 1.2~R$_\odot$ based on a sample of four active regions.}

{However, plasma emission is the main emission mechanism at low radio frequencies ($<$100~MHz) as can be seen from Figure 4, since the plasma frequency is dominant. Emission at the electron-cyclotron frequency at 100~MHz would only occur at very low heights and it would not escape the overlying plasma layers due to free-free absorption [Bastian et al. 1998]. type III radio bursts and solar S bursts are most likely emitted by the plasma emission mechanism.}

\section{Conclusion}

{The characteristics of over 800 type III radio bursts and 3000 solar S bursts were presented in this study using high spectral and temporal resolution dynamic spectra from LOFAR tied-array beam observations. type III radio bursts and solar S bursts are morphologically different bursts which seem to always occur concurrently in the presence of active regions on the Sun. }

{S bursts have a drift rate that is 1/2 that of type III bursts. The drift rate of both these bursts increases with frequencies in a similar manner. These bursts are therefore generated at the plasma frequency since plasma emission is widely accepted as the emission for type III radio bursts. The varying drift rates of these bursts suggest that the electrons that generated the bursts belonged to different populations, most likely originating from different spatial locations which is confirmed by tied-array beam observations (Figure 1). The drift rates of S-bursts and type IIIs obtained in this study are systematically higher than those obtained in previous studies by McConnell [1982], Mann et al. [1999] and Melnik et al. [2010]. In the case of type IIIs, Mann et al. [1999] analysed only a small set of bursts. On the other hand, Alvarez and Haddock [1973] used observations covering a complete 11-year solar cycle and the Type III drift rates obtained compare very well to this study (the exponent of the fit is almost the same as the LOFAR fit in Figure 2 within error). In the case of S-bursts, McConnell [1982]  observed only a few S bursts and Melnik et al. [2010] observed a significant number of S bursts but over a short bandwidth (10--30~MHz). More studies over large bandwidths are necessary to constrain the drift rate dependence coefficients for S bursts in Figure 2.}

{The drift rate of both solar S bursts and type III radio bursts is indeed different to that of jovian S bursts, which are a well known example of ECM emission and electrons travelling adiabatically along the Io-Jupiter flux tubes. Plasma emission was shown to be the main emission mechanism on the Sun at low radio frequencies ($<$100~MHz) where the plasma frequency is dominant. Solar S bursts are most likely emitted by the plasma emission mechanism, as type III radio bursts. Future high resolution imaging observations are necessary to determine the origin of solar S bursts bursts and to analyse how they compare spatially to the concurrent type III radio bursts.  }

\section*{References}
\everypar={\hangindent=1truecm \hangafter=1}

Abranin,~E.\,P., L.\,L.~Bazelyan, and Y.\,G.~Tsybko, Stability of the
parameters of decameter type-III bursts over the 11-year solar
activity cycle - Rates of frequency drift of radio burst, \textsl{Soviet Astron.}, \textbf{34}, 74--78, 1990.

Alvarez,~H., and F.\,T.~Haddock, Solar wind density model from km-wave
Type III bursts, \textsl{Solar Phys.}, \textbf{29}, 197-209, 1973.

Bastian,~T.\,S., A.\,O.~Benz, and D.\,E.~Gary, Radio emission from
solar flares, \textsl{Ann. Rev. Astron. Astrophys.}, \textbf{36}, 131, 1998.

Boischot,~A., Caract\`eres d'un type d'\'emission hertzienne associ\'e \`a certaines \'eruptions chromosph\'eriques, \textsl{Comptes Rendu Acad. Sci.}, \textbf{244}, 1326, 1957.

Briand,~C., A.~Zaslavsky, M.~Maksimovic, P.~Zarka, A.~Lecacheux, H.\,O.~Rucker, A.\,A.~Konovalenko, E.\,P.~Abranin, V.\,V.~Dorovskyy, A.\,A.~Stanislavsky, and V.\,N.~Melnik, Faint solar radio structures from decametric observations, \textsl{Astron. Astrophys.}, \textbf{490}, 339--344, 2008.

Dulk, G. A., 1985, Radio emission from the sun and stars, \textit{Annual review of astronomy and astrophysics}, Volume 23 (A86-14507 04-90). Palo Alto, CA, Annual Reviews, Inc., p. 169-224, 1985.

Dulk,~G.\,A., Y.~Leblanc, T.\,S.~Bastian, and J.-L.~Bougeret, Acceleration of electrons at type II shock fronts and production of shock-accelerated type III bursts, \JGR{105}{27343-27352}{2000}

Ellis,~G.\,R.\,A., Fine structure in the spectra of solar radio bursts. \textsl{Aus. J. Phys.}, \textbf{22}, 177, 1969.

Ellis,~G.\,R.\,A., Observations of the Jupiter S--bursts between 3.2 and 32~MHz, \textsl{Aus. J. Phys.}, \textbf{35}, 165--175, 1982.

Hess,~S.\,L.\,G., F.~Mottez, and P.~Zarka, Jovian S burst generation by Alfv\'en waves, \textsl{J. Geophys. Res.}, \textbf{112}, 11, 11212, 2007.

Hess,~S.\,L.\,G., F.~Mottez, and P.~Zarka, Effect of electric potential structures on Jovian S-burst morphology, \textsl{Geophys. Res. Lett.}, \textbf{36}, 14101, 2009.

Krupar,~V., M.~Maksimovic, O.~Santolik, B.~Cecconi, Q.\,N.~Nguyen, S.~Hoang, and K.~Goetz, The apparent source size of type III radio bursts: Preliminary results by the STEREO/WAVES instruments, 12th Int. Solar wind conference, \textsl{AIP Conf. Proc.}, \textbf{1216}, 284--287, 2010.

Lin, R. P., Non-relativistic Solar Electrons, \textit{Space Science Reviews}, Volume 16, Issue 1-2, pp. 189-256, 1974.

Ma, Y.; Xie, R.; Zheng, X.; Wang, M; Yi-hua, Y., A Statistical Analysis of the type-III Bursts in Centimeter and Decimeter Wavebands, \textit{Chinese Astronomy and Astrophysics}, Volume 36, Issue 2, p. 175-186, 2012.

Mann,~G., F.~Jansen, R.\,J.~MacDowall, M.\,L.~Kaiser, and R.\,G.~Stone, A heliospheric density model and type III bursts, \textsl{Astron. Astrophys.}, \textbf{348}, 614--621, 1999.

Mann, G.; Klassen, A., Shock accelerated electron beams in the solar corona, In: \textit{Solar variability: from core to outer frontiers.} The 10th European Solar Physics Meeting, 9 - 14 September 2002, Prague, Czech Republic. Ed. A. Wilson. ESA SP-506, Vol. 1. Noordwijk: ESA Publications Division, ISBN 92-9092-816-6, 2002, p. 245 - 248, 2002.

McConnell, D., Spectral characteristics of solar S bursts, \textit{Solar Physics}, 78, 253, 1982.

Melnik,~V.\,N., A.\,A.~Konovalenko, H.\,O.~Rucker, V.\,V.~Dorovskyy, E.\,P.~Abranin, A.~Lecacheux, and A.\,S.~Lonskaya, Solar S-bursts at frequencies of 10--30~MHz, \textsl{Solar Phys.}, \textbf{264}, 103-117, 2010.

Melrose,~D.\,B., Collective plasma radiation processes, \textsl{Ann. Rev. Astron. Astrophys.}, \textbf{29}, 31, 1991.

Morosan,~D.\,E. et al. (84 co--authors), LOFAR tied--array imaging of type III solar radio bursts, \textsl{Astron. Astrophys.}, \textbf{568}, id.A67, 8 pp., 2014.

Morosan,~D.\,E. et al. (52 co--authors), LOFAR tied--array imaging and spectroscopy of solar S bursts, \textsl{Astron. Astrophys.}, \textbf{580}, id.A65, 6 pp., 2015.

Morosan,~D.\,E., P.~Zucca, D.\,S.~Bloomfield, and P.\,T.~Gallagher, Conditions for electron--cyclotron maser emission in the solar corona, \textsl{Astron. Astrophys.}, \textbf{589}, id.L8, 4 pp., 2016.

R\'egnier, S., A new approach to the maser emission in the solar corona, \textit{Astron. Astrophys.}, Volume 581, id.A9, 9 pp., 2015

Robinson, P. A.; Willes, A. J.; Cairns, I. H., \textit{Dynamics of Langmuir and ion-sound waves in type III solar radio sources}, Astrophysical Journal, Part 1 (ISSN 0004-637X), vol. 408, no. 2, p. 720-734, 1993.

Saint-Hilaire, P.; Vilmer, N.; Kerdraon, A., A Decade of Solar type III Radio Bursts Observed by the Nancay Radioheliograph 1998-2008, \textit{The Astrophysical Journal}, Volume 762, Issue 1, article id. 60, 16 pp, 2013.

Schrijver, C. J.; De Rosa, M. L., Photospheric and heliospheric magnetic fields, \textit{Solar Physics}, v. 212, Issue 1, p. 165-200, 2003.

Stappers,~B.\,W., et al. (93 co-authors), Observing pulsars and fast transients with LOFAR, \textsl{Astron. Astrophys.}, \textbf{530}, A80, 2011.

van~Haarlem,~M., et al. (200 co--authors), LOFAR: The LOw-Frequency ARray, \textsl{Astron. Astrophys.}, \textbf{556}, id.A2, 53 pp., 2013.

Wild, J. P., and L. L. McCready. Observations of the Spectrum of High-Intensity Solar Radiation at Metre Wavelengths. I. The Apparatus and Spectral types of Solar Burst Observed. \textsl{Aust. J. Sci. Res. A}, \textbf{3}, 387, 1950.

Wild, J. P. Observations of the Spectrum of High-Intensity Solar Radiation at Metre Wavelengths. II. Outbursts. \textsl{Aust. J. Sci. Res. A}, \textbf{3}, 399, 1950.

Wild,~J.\,P., Observations of the spectrum of high--intensity solar radiation at metre wavelengths. III. Isolated bursts, \textsl{Aust. J. Sci. Res. A}, \textbf{3}, 541--557, 1950.

Wild,~J.\,P., K.\,V.~Sheridan, and G.\,H.~Trent, Transverse motion of the source of solar radio bursts, in \textsl{Proc. IAU/URSI Symp. Paris Symposium on Radio Astronomy}, edited by R.\,N.~Bracewell, Stanford Univ. Press, Stanford, USA, p.176, 1959.

Zarka, P.; Farges, T.; Ryabov, B. P.; Abada-Simon, M.; Denis, L., \textit{A scenario for Jovian S-bursts}, Geophysical Research Letters, Volume 23, Issue 2, p. 125-128, 1996.

Zeitsev, V. V.; Zlotnik, E. Y., A Mechanism for Generating the Solar S-Burst Trains, \textit{Soviet Astronomy Letters}, vol. 12, Mar.-Apr. 1986, p. 128-131. Translation Pisma v Astronomicheskii Zhurnal, vol. 12, Apr. 1986, p. 311-317, 1986.

Zucca, P., Carley, E.~P., Bloomfield, D.~S., \& Gallagher, P.~T.\ , The formation heights of coronal shocks from 2D density and Alfvén speed maps. \textsl{Astron. Astrophys.}, 564, A47, 2014.

\end{document}